\documentstyle[]{ioplppt}  
\begin{document}              
\jl{6}
\title{A new test of conservation laws and Lorentz invariance in 
relativistic gravity}
\author{J F Bell\dag~and T Damour\S}
\address{\dag University of Manchester, NRAL, Jodrell
       Bank, Macclesfield, Cheshire SK11~9DL, UK, email: jb@jb.man.ac.uk}
\address{\S Institut des Hautes Etudes Scientifiques, F-91440
Bures-sur-Yvette, France, and DARC, CNRS-Observatoire de Paris, 92195 Meudon,
France, email: damour@ihes.fr}

\begin{abstract}               
General relativity predicts that energy and momentum conservation laws hold
and that preferred frames do not exist. The parametrised post-Newtonian
formalism (PPN) phenomenologically quantifies possible deviations from
general relativity.  The PPN parameter $\alpha_{3}$ (which identically
vanishes in general relativity) plays a dual role in that it is associated
both with a violation of the momentum conservation law, and with the
existence of a preferred frame.  By considering the effects of $\alpha_{3}
\neq 0$ in certain binary pulsar systems, it is shown that $|\alpha_{3}| <
2.2 \times 10^{-20}$ (90\% CL). This limit improves on previous results by
several orders of magnitude, and shows that pulsar tests of $\alpha_{3}$
rank (together with Hughes-Drever-type tests of local Lorentz invariance)
among the most precise null experiments of physics.
\end{abstract}
\section{Introduction}            

Conservation laws have long been a foundation of physics and careful tests
of their validity have led to important discoveries, such as neutrinos.
Most theories of gravity, starting with general relativity, are based on an
invariant action principle and therefore predict that the energy and
momentum of isolated gravitating systems are conserved. However, direct
experimental constraints on those conservation laws are far less extensive
than for the other fundamental forces of nature.  Another cornerstone of
physics is the absence of preferred frames in local experiments (Local
Lorentz Invariance).  This property may be violated in certain theories of
gravity allowing for the existence of long-range vector fields.

The parametrised post-Newtonian formalism (PPN) phenomenologically
quantifies possible deviations from general relativity. Such deviations are
measured by a set of ten parameters:$\gamma -1, \beta -1, \xi, \alpha_{1},
\alpha_{2}, \alpha_{3}, \zeta_{1}, \zeta_{2}, \zeta_{3}, \zeta_{4}$ \cite{wil93}.
To each of these parameters is associated a class of observable,
non-Einsteinian effects. The parameter $\alpha_{3}$ plays a special role in
that it is associated both with a violation of the momentum conservation law,
and with the existence of preferred frames ($\alpha_{3} \equiv 0$ in general
relativity).

The most striking observable effect induced by a putative $\alpha_{3}
\neq 0$ is the existence of a non-zero self-acceleration for a rotating body,
perpendicular to its spin axis and absolute velocity
\cite{nw72,n73,wil93}. As a consequence, the total acceleration of a body,
which is a member of a gravitating system, can be decomposed as
\begin{equation}
{\bf a} = {\bf a}_{\rm Newt} + {\bf a}_{\rm nbody} + {\bf a}_{\alpha_{3}} 
\label{e:acc}
\end{equation}
where ${\bf a}_{\rm Newt}$ is a two-body Newtonian-like $1/R^2$ force
(modified by possible equivalence-principle-violation effects), $ {\bf
a}_{\rm nbody}$ denotes relativistic n-body effects, and where $ {\bf
a}_{\alpha_{3}}$ is the self-acceleration resulting from the violation of
momentum conservation associated with $\alpha_{3} \neq 0$
\cite{nw72,n73,wil93}.  For a nearly spherical body, rotating with angular
velocity ${\bf \Omega}$,
\begin{equation}
{\bf a}_{A\alpha_{3}} = -{1 \over 3} \alpha_{3} {E_{A}^{\rm grav} \over
m_{A}c^{2}} \left({\bf w} + {\bf v} \right) \times {\bf \Omega}
\label{e:aself}
\end{equation}
where $E_{A}^{\rm grav}$ is the gravitational self energy of body $A$ having mass
$m_{A}$, moving with  velocity ${\bf w}+ {\bf v}$, with respect to the
absolute rest frame. For later convenience, we have decomposed 
this velocity into the absolute velocity $\bf w$ of the centre of mass
of the considered system, and the peculiar velocity $\bf v$ 
of body $A$ with respect to the centre of mass frame.
Instead of the gravitational self energy $E_A$, one can introduce the 
(dimensionless) ``compactness'' parameter 
$c_A = -2 \partial {\rm ln} m_A / \partial {\rm ln} G$
which is approximatively given by
 $c_{A} = - 2 E_{A}^{\rm grav}/m_{A}c^{2}$ \cite{de92}.
We see from equation (\ref{e:aself}) that, given
an absolute velocity ${\bf w}$, $\Omega c_{A} = 2 \pi c_{A} /
P$ defines a figure of merit for the selection of  bodies testing $\alpha_3$. 
For the sun $c_{A} \sim 10^{-6}$, and for a white dwarf
$c_{A} \sim 10^{-4}$. By contrast, neutron stars have compactness 
parameters of order unity.
Considering a range of equations of state for neutron stars, 
Ref. \cite{de92} found that $c_A$ takes a
median value of $ 0.21 m_A/M_{\odot}$.
Therefore, rapidly rotating neutron stars are, by a very large 
factor, the best objects for constraining $\alpha_{3}$. In the 
present paper, we shall show that existing data on certain 
long-period, quasi-circular binary pulsar systems allow one 
to constrain $\alpha_{3}$ at the $10^{- 20}$ level.

\section{Previous Limits}

As shown by Nordtvedt and Will \cite{nw72,wil93}, a non-zero
$\alpha_{3}$ induces a contribution to the perihelion precession of the
planets in the solar system. The two planets with the best measurements of
periastron advance were Earth and Mercury. By combining the observations for
two planets it is possible to eliminate the terms involving the well-known
Eddington parameters $\gamma$ and $\beta$, obtaining $|49 \alpha_{1} -
\alpha_{2} - 6.3 \times 10^{5} \alpha_{3} - 2.2\xi| < 0.1$
\cite{wil93}. Using the limits on other PPN parameters a limit of
$|\alpha_{3}| < 2 \times 10^{-7}$ was thus obtained. 

A tighter limit on $\alpha_{3}$ has been obtained by considering the effect
of the acceleration (\ref{e:aself}) on the observed pulse periods of
isolated pulsars.  The observed pulse period $P = P_{0}(1 + v_{r}/c)$,
contains contributions from both the intrinsic pulse period $P_{0}$ and the
Doppler effect resulting from the radial velocity $v_{r}$.  Consequently any
radial acceleration $a_{r}$ contributes to the observed period derivative
$\dot{P}$,
\begin{equation}
{\dot{P} \over P} = {\dot{P}_{0} \over P} + {a_{r} \over c}
\label{e:pdot}
\end{equation}

Self accelerations are directed perpendicular to both the spin axis and
the absolute velocity of the spinning body. 
If self accelerations were contributing
strongly to the observed period derivatives of pulsars, roughly equal numbers
of positive and negative observed period derivatives would be expected,
since the spin axes and therefore the self accelerations are randomly
oriented.  However the observed distribution (excluding those pulsars in
globular clusters) contains only positive period derivatives, allowing a
limit to be placed on $\alpha_{3}$. This was done initially using normal
pulsars \cite{wil93} to obtain a limit of $|\alpha_{3}| < 2 \times 10^{-10}$
and more recently using millisecond pulsars \cite{bel96} to obtain a limit
of $|\alpha_{3}| < 5 \times 10^{-16}$. The later case incorrectly included
binary pulsars in the sample. [As will be shown in the following section the
effects of self accelerations in a binary system are more complicated.]
Nevertheless, if one restricts oneself to a sample of
isolated millisecond pulsars,  a limit
quantitatively similar to that above, though slightly less stringent,
 can be obtained.

\section{Polarisation of Binary Orbits}
\label{s:pol}

When considering a binary system there is a perturbing self acceleration
felt by each body. These self accelerations perturb both the centre of mass
motion of the system, and the relative orbital dynamics. We focus on the
perturbations of the relative motion which turn out to be a much more sensitive
probe of $\alpha_3$. The perturbation of the relative acceleration ${\bf
a}_{\alpha_3} = {\bf a}_{A\alpha_3} - {\bf a}_{B\alpha_3}$, where $A$ labels
the pulsar and $B$ its companion, can be written as
\begin{equation} \fl
{\bf a}_{\alpha_{3}} = {\bf a}_{\alpha_{3}}^{\rm Stark} + 
{\bf a}_{\alpha_{3}}^{\rm Zeeman}
\label{e:st_zee}
\end{equation}
where
\begin{equation}
{\bf a}_{\alpha_{3}}^{\rm Stark} =
{\alpha_{3} \over 6} {\bf w} \times (c_{A} {\bf \Omega}_{A} - c_{B}
{\bf \Omega}_{B}) \, ,
\label{e:stark}
\end{equation}
\begin{equation}
{\bf a}_{\alpha_{3}}^{\rm Zeeman} = {\alpha_{3} \over 6} {\bf v} \times
(x_{B} c_{A} {\bf \Omega}_{A} + x_{A} c_{B} {\bf \Omega}_{B}) 
\label{e:zeeman}
\end{equation}
Here $c_{A}, \Omega_{A}, x_{A} = m_{A}/(m_{A} + m_{B})$ are the
compactness, angular velocity and mass fraction for the pulsar. The names given
to the perturbing accelerations (\ref{e:stark}) and (\ref{e:zeeman}) have been
chosen by analogy with the well-known effects of constant external electric and
magnetic fields on the classical dynamics of an atom. These two perturbations
have very distinct effects on the relative orbital dynamics.

By a generalisation of Larmor's theorem, the ``Zeeman'' contribution
(\ref{e:zeeman}) is easily seen to cause a slow overall precession of the orbit
with angular velocity
\begin{equation}
{\bf \Omega}_{\rm precession} = -{\alpha_3 \over 12} (x_B c_A {\bf \Omega}_A +
x_A c_B {\bf \Omega}_B) \, . \label{e:precession}
\end{equation}
The theory of the formation of binary pulsars leads us to expect that, to a
first approximation, the spin vectors ${\bf \Omega}_A$, ${\bf \Omega}_B$ are
parallel to the orbital angular momentum. This means that the effects of the
perturbation (\ref{e:precession}) give rise to non-observable, very small
additional contributions to the orbital period and the periastron precession.

By contrast, the ``Stark'' contribution (\ref{e:stark}), which represents a
constant perturbing acceleration, leads to a forced eccentricity, polarising
the orbit along a fixed direction. Such a ``Stark'' polarisation (which is the 
DC analogue of the Nordtvedt polarisation of the lunar orbit \cite{nor68b})
has already been studied in certain binary pulsars when considering the effect
of a violation of the strong equivalence principle \cite{ds91}. The best
systems for constraining the polarisation effects caused by (\ref{e:stark}) are
the long-orbital-period, quasi-circular binary millisecond pulsars with white
dwarf companions. For a neutron star and white dwarf $c_B \ll c_A$ and
$\Omega_B \ll \Omega_A$, so that equation (\ref{e:stark}) reduces to
\begin{equation}
{\bf a}_{\alpha_{3}}^{\rm Stark} \simeq {\alpha_{3} \over 6} c_A {\bf
w} \times {\bf \Omega}_{A} \, . \label{e:starkA}
\end{equation}

For this acceleration to be in a fixed direction, ${\bf w}$
and ${\bf \Omega}_{A}$ must remain fixed in space. Any precession of the
pulsar spin axes will be very slow, if it occurs at all. The absolute
velocity of pulsars in our galactic neighbourhood is taken to be $\vert {\bf
w}\vert = 369$~km~s$^{-1}$, as determined from observation of the cosmic
microwave background \cite{fcc+94}. Some recent results \cite{lp94} question the
validity of the cosmic microwave background as an absolute reference frame. As a
sample of pulsars are used, only the magnitude of ${\bf w}$ is important and
this is similar in both cases.

It is also required that the centre of mass of the binary system does not
move appreciably in the Galaxy during the build up of the polarisation
induced by (\ref{e:starkA}), otherwise ${\bf w}$ will vary and the
polarisation force (\ref{e:starkA}) must be decomposed in a sum of
monochromatic terms. This is equivalent to a corresponding requirement for
the test of the strong equivalence principle \cite{ds91}, ie
\begin{equation}
{\omega_{\rm Gal} \over \omega_{R}} =
\left({ P_{b} \over 1364({\rm days})} \right)^{5/3}
\left({M \over 1.7 M_{\odot}} \right)^{-2/3} \ll 1
\label{e:wgal}
\end{equation}
where $\omega_{\rm Gal}$ is the angular rotation rate of the Galaxy near the
sun, $\omega_{R}$ is the relativistic periastron advance and $M = m_{A} +
m_{B}$.

The equation of motion for a binary system with the perturbation
(\ref{e:stark}) is very similar to the equation of motion for the strong
equivalence principle test \cite{ds91}
\begin{equation}
{{\rm d}^{2}{\bf r} \over {\rm d}t^{2}} + {G M{\bf r} \over r^{3}} =
{\bf R} + {\bf a}_{\alpha_{3}}^{\rm Stark}
\label{e:mot}
\end{equation}
where ${\bf R}$ contains the relativistic forces responsible for
periastron precession. We can use the results of \cite{ds91}. Considering
systems for which $e \ll 1$ while letting ${\bf w} \times {\bf
\Omega}_{A} = w \Omega_{A} \sin \beta {\bf e}_{u}$ (where ${\bf e}_u$ is a unit
vector) and $\Omega_{A} = 2 \pi / P$ gives the solution
\begin{equation}
{\bf e}(t) = {\bf e}_{R}(t) + {\bf e}_{\alpha_{3}} = 
{\bf e}_{R}(t) + 
\alpha_{3} {c_{A} w \over 24 \pi} {P_{b}^{2} \over P } {c^{2} \over G M}
\sin \beta {\bf e}_{u} \, .
\label{e:ecc}
\end{equation}
Here ${\bf e} (t)$ is the total, observable eccentricity vector, ${\bf
e}_{R}(t)$ is the intrinsic eccentricity vector, which rotates with angular
frequency $\omega_{R}$ due to relativistic periastron precession and ${\bf
e}_{\alpha_{3}}$ is the forced eccentricity vector caused by $\alpha_3 \not=
0$. It is assumed that since considerable mass transfer has taken place,
the orbital and spin angular momenta are aligned, which results in the forcing
term (\ref{e:stark}) due to $\alpha_{3}$ being parallel to the orbital plane.
Given an observed eccentricity $e_{\rm obs} = \vert {\bf e}(t) \vert$ measured
at some time $t$, equation (\ref{e:ecc}) may be used to place a limit on
$\alpha_{3}$. From equation (\ref{e:ecc}) it can be seen that 
\begin{equation}
P_{b}^{2} / e P
\label{e:fom}
\end{equation} 
defines a figure of merit for selecting which binary pulsar systems will
provide the best limit. This figure of merit differs from the one $(P_b^2 /e)$
selecting the best tests of the strong equivalence principle.

\section{Confidence level of the limit}

Unfortunately, in equation (\ref{e:ecc}) there are two parameters which are
not constrained by the observations, $\beta$ and ${\bf
e}_{R}(t)$. Firstly, the angle $\beta$ may be very small causing
$\alpha_{3}$ effects to contribute weakly to the observed
eccentricity. Secondly, there is a finite probability that ${\bf
e}_{R}(t)$ might have cancelled most of the term due to $\alpha_{3}$ leaving
only a small observed eccentricity. In order to impose a limit on $\alpha_{3}$ at
a particular confidence level a quantitative assessment of the probability of
these independent effects is needed. 

The problem of ${\bf e}_{R}(t)$ potentially cancelling a second term was
first considered by Damour and Sch\"afer \cite{ds91} who decomposed the
problem into three cases. When ${\bf e}_{R}(t) \gg {\bf e}_{\alpha_{3}}$ or
${\bf e}_{R}(t) \ll {\bf e}_{\alpha_{3}}$ it is sufficient to use the
observed eccentricity $e_{\rm obs}$ for obtaining a limit. When ${\bf e}_{R}(t)
\simeq {\bf e}_{\alpha_{3}}$ and the angle between ${\bf e}_{R}(t)$ and the
opposite of ${\bf e}_{\alpha_{3}}$ is $\theta$, then $e_{\rm obs} =
e_{\alpha_3} 2 \vert \sin (\theta /2)\vert$ was used. A slightly more
precise approach was taken by Wex \cite{wex96} who defined a function
\begin{equation}
S(\theta) = \left\{
\begin{array}{ll}
|\sin \theta | & {\rm if}~ | \theta | < \pi/2 \\ 1 & {\rm if}~ | \theta | \geq
\pi/2 \end{array} \right\} \, .
\label{e:sth}
\end{equation}
Here the angle $\theta$ between ${\bf e}_{R}(t)$ and
$-{\bf e}_{\alpha_{3}}$ is taken to vary between $-\pi$ and $\pi$. Given
the magnitude of the forced eccentricity $e_{\alpha_3} = \vert {\bf
e}_{\alpha_3} \vert$ and the value of the angle $\theta$, it is easily seen
that when the magnitude of the intrinsic eccentricity $\vert {\bf e}_R \vert$
is allowed to vary one can write the inequality
\begin{equation}
e_{\rm obs} \geq e_{\alpha_3} \ S(\theta) \, . \label{e:ineq}
\end{equation}
[Indeed, when $\vert \theta \vert < \pi /2$ the worst cancellation arises when
the resulting eccentricity vector ${\bf e} = {\bf e}_R + {\bf e}_{\alpha_3}$ is
perpendicular to ${\bf e}_R$, while when $\vert \theta \vert > \pi /2$ the
worst case is $\vert {\bf e}_R \vert =0$.] Using the inequality (\ref{e:ineq})
and rearranging equation (\ref{e:ecc}) gives the upper bound
\begin{equation}
| \alpha_{3} | \leq {24 \pi \over c_{A} w } {e P \over P_{b}^{2}} {GM \over
c^{2}} {1 \over S(\theta) \sin \beta} \equiv {K_{i} \over S(\theta) \sin \beta}
\label{e:lim}
\end{equation}
Given the quantity $K_{i}$ determined from observed parameters, the
probability that $|\alpha_{3} |$ is {\it greater} than some number $\alpha$ is
{\it smaller} than 
\begin{equation}
F(K_{i}/\alpha) = {\rm Prob}[S(\theta) \sin \beta <
K_{i}/\alpha] \, .
\end{equation}

For sufficiently old systems, ${\bf e}_R (t)$ has had the time to make many
turns and we can consider that the orbital phase $\theta$ is randomly
distributed over the interval $[-\pi ,\pi]$. [The pulsars included in our
sample below have old white dwarf companions, confirming that this oldness
condition is satisfied.] Given ${\bf w}$, the spin vector ${\bf \Omega}_A$
can point in an arbitrary direction on the unit sphere. Therefore the polar
angle $\beta$ is distributed with the probability law ${1\over 2} \sin \beta
d\beta$. In other words, the quantity $c_{\beta} = \cos \beta$ is
distributed uniformly over $[-1,1]$. Hence, since the distributions are
symmetric about $\theta = 0$ and $c_{\beta} = 0$,
\begin{equation} \fl
F\left({K_{i} \over \alpha }\right) = {\rm Prob} \left[S(\theta)\sqrt{1 -
c_{\beta}^{2}} < {K_{i} \over \alpha}\right] ^{|\theta| < \pi}_{|c_{\beta}|
< 1} = 4 {\rm Prob} \left[S(\theta)\sqrt{1 - c_{\beta}^{2}} < {K_{i} \over
\alpha}\right] ^{0 < \theta < \pi}_{0 < c_{\beta} < 1} \\
\label{e:prob}
\end{equation}
\begin{equation} \fl
F\left({K_{i} \over \alpha }\right) = 1 - \sqrt{1 - \left({K_{i} \over
 \alpha }\right)^{2}} + {1 \over \pi}
\int_{0}^{\sqrt{1 - (K_{i} / \alpha )^{2}}}
dc_{\beta} \arcsin \left[ {K_{i} / \alpha \over \sqrt {1 - c_{\beta}^{2}
}}\right]
\label{e:prob2}
\end{equation}
Probabilities evaluated from this function using the {\tt Mathematica} package to
numerically integrate the second term are shown in Table \ref{t:prob} for a
range of values of $K_{i}/ \alpha$.

\begin{table}
\caption{The probability that $S(\theta) \sin \beta < K_{i}/ \alpha$}
\begin{indented}
\item[]\begin{tabular}{cc} \br
$K_{i}/\alpha$ & Probability \\ \mr
0.00 & 0.000 \\
0.10 & 0.045 \\
0.20 & 0.092 \\
0.30 & 0.143 \\
0.40 & 0.201 \\
0.50 & 0.267 \\
0.60 & 0.343 \\
0.70 & 0.432 \\
0.80 & 0.540 \\
0.85 & 0.604 \\
0.90 & 0.681 \\
0.93 & 0.736 \\
0.96 & 0.804 \\
0.97 & 0.831 \\
0.98 & 0.864 \\
0.99 & 0.905 \\
1.00 & 1.000 \\ \br
\end{tabular}
\label{t:prob}
\end{indented}
\end{table}

Since the observational results for pulsars are independent, the combined
probability, given the existence of several pulsars with small values for
$K_{i}$, satisfies the inequality
\begin{equation}
 1 - ({\rm CL}/100) \equiv {\rm Prob} [ |\alpha_{3}| > \alpha ] < \prod_{i} F(
K_{i} / \alpha) \, . \label{e:prod}
\end{equation}
Here, we have introduced the confidence level (CL) with which one can reject
the hypothesis that $\vert \alpha_3 \vert$ is larger than some given $\alpha$.

When the product on the right-hand side is equal to $0.1$, $|\alpha_{3}| <
\alpha$ at {\it better} than the 90\% confidence level. Note that when
$K_{i} \geq \alpha$, $F(K_{i}/ \alpha) = 1$, so that including such a pulsar
does not improve the limit.

\section{The Limit}

In the most recent compilation of fast pulsars in binary systems
\cite{cam96}, there are many systems which satisfy the constraints discussed
in the previous two sections. Using the result of \cite{de92} for the median
value of $c_A$ quoted above, one finds $K_i = 1.925 \times 10^{-13} (1+m_B /
m_A) eP / P_b^2$ where the pulsar period $P$ is expressed in milliseconds and
the orbital period $P_b$ in days. After evaluating $K_{i}$ for each of those
binary pulsars, the objects with the smallest $K_{i}$ were chosen. They are
listed in Table \ref{t:lim}. The probability that $S(\theta) \sin \beta <
K_{i}/\alpha$ was evaluated using equation (\ref{e:prob2}) for several values of
$\alpha$ which are shown in Table \ref{t:lim}. The combined probabilities and
confidence levels from using several pulsars were
then determined using equation (\ref{e:prod}) (See Table \ref{t:lim}).

\begin{table}
\caption{$F(K_{i}/ \alpha)$ for the test systems. $K_{i}$ and $\alpha$ are in units of $10^{-20}$.}
\begin{indented}
\item[]\begin{tabular}{l|c|ccccc} \br
Pulsar       & $K_{i}$ & $\alpha = 3.7$ & $\alpha = 2.9$ & $\alpha = 2.4$ & $\alpha = 2.2$ & $\alpha = 1.95$ \\ \mr
B1855+09     &  17.6   & 1.00 & 1.00 & 1.00 & 1.00 & 1.00 \\
J2317+1439   &  12.6   & 1.00 & 1.00 & 1.00 & 1.00 & 1.00 \\
J1455$-$3330 &  5.45   & 1.00 & 1.00 & 1.00 & 1.00 & 1.00 \\
B1953+29     &  3.24   & 0.64 & 1.00 & 1.00 & 1.00 & 1.00 \\
J1643$-$1224 &  2.30   & 0.36 & 0.54 & 0.80 & 1.00 & 1.00 \\
J1640+2224   &  1.91   & 0.28 & 0.40 & 0.54 & 0.62 & 0.86 \\
J2229+2643   &  1.83   & 0.27 & 0.37 & 0.50 & 0.58 & 0.76 \\
J2019+2425   &  1.79   & 0.26 & 0.36 & 0.48 & 0.55 & 0.71 \\
J1713+0747   &  1.76   & 0.25 & 0.35 & 0.47 & 0.54 & 0.68 \\ \mr
$\prod_{i} F( K_{i} / \alpha) $ 
             &         & 0.001  & 0.01 & 0.05 & 0.10 & 0.32 \\ \mr
CL $>$       &         & 99.9\% & 99\% & 95\% & 90\% & 68\% \\ \br
\end{tabular}
\label{t:lim}
\end{indented}
\end{table}

Using, for instance, the 90\% confidence level gives a limit $|\alpha_{3}| < 2.2
\times 10^{-20}$. This is more than a factor of $10^{4}$ better than the
previous limit \cite{bel96}, and by far the tightest limit on any of the PPN
parameters. 

The only other ultra-high-precision null experiments giving limits of order
$10^{-20}$ on a dimensionless theoretical parameter we are aware of are some
recent Hughes-Drever-type tests of local Lorentz invariance
\cite{pre85,lam86,chu89}. Figure 14.2 of \cite{wil93} shows the limits on
the parameter $\delta = c_0^2 / c_e^2 -1$, where $c_0$ is the limiting speed
of massive particles, and $c_e$ the speed of light. It is remarkable that
tests involving binary pulsars can rank, together with modern laser-cooled
trapped atom experiments, among the most precise null experiments of
physics.

\section*{Acknowledgments}

J.F.B. thanks the Institut des Hautes Etudes Scientifiques for its hospitality
during the conception of this work.


\section*{References}

\begin{thebibliography}{8}

\bibitem{wil93}
Will, C.~M. 1993, { Theory and Experiment in Gravitational Physics},
  (Cambridge: Cambridge University Press)

\bibitem{nw72} Nordtvedt, K. \& Will, C.M. 1972, ApJ, 177, 775

\bibitem{n73} Nordtvedt, K. 1973, Phys. Rev. D, 7, 2347

\bibitem{de92}
Damour, T. \& Esposito-Far\`ese, G. 1992, Class. Quant. Grav., 9, 2093

\bibitem{bel96}
Bell, J.~F. 1996, ApJ, 462, 287

\bibitem{nor68b} Nordtvedt, K. 1968, Phys. Rev., 170, 1186

\bibitem{ds91}
Damour, T. \& Sch\"afer, G. 1991, Phys. Rev. Lett., 66, 2549

\bibitem{fcc+94}
Fixsen, D.~J. {et al.}  1994, ApJ, 420, 445

\bibitem{lp94}
Lauer, T.~R. \& Postman, M. 1994, ApJ, 425, 418

\bibitem{wex96}
Wex, N. 1996, A\&A, In Press

\bibitem{cam96}
Camilo, F. 1996, in { High Sensitivity Radio Astronomy}, ed.\ N. Jackson,
  Cambridge University Press, In Press

\bibitem{pre85}
Prestage, J.D. et al. 1985, Phys. Rev. Lett., 54, 2387 

\bibitem{lam86}
Lamoreaux, S.K. et al. 1986, Phys. Rev. Lett., 57, 3125 

\bibitem{chu89}
Chupp, T.E. et al. 1989, Phys. Rev. Lett., 63, 1541

\end{thebibliography}

\end{document}